# Physical properties, strontium ordering and structural modulation in layered hexagonal $Sr_{0.35}CoO_2$

H.X. Yang*, Y.G. Shi, Y.Q. Guo, X. Liu, R.J. Xiao, J.L. Luo and J.Q. Li

Beijing National Laboratory for Condensed Matter Physics, Institute of Physics, Chinese Academy of Sciences, Beijing 100080, China

Layered $Sr_{0.35}CoO_2$ has been synthesized by means of an ion exchange reaction from $Na_{0.7}CoO_2$. Resistivity measurements show that this material can be either metallic or semiconducting depending on the annealing conditions. Magnetic susceptibility, following the Curie-Weiss law, increases with lowering temperature and shows a small anomalous kink at around 30 K. Transmission-electron-microscopy observations reveal the presence of two superstructures arising respectively from the intercalated Sr-ordering (a compositional modulation) with $\mathbf{q}_1 = \mathbf{a}^* / 3 + \mathbf{b}^* / 3$ and a periodic structural distortion (a transverse structural modulation) with $\mathbf{q}_2 = \mathbf{a}^* / 2$.

Author to whom correspondence should be addressed: hxyang@blem.ac.cn

Layered deintercalatable alkali metal oxides, such as $Li_xCoO_2$ ($0.5 \leq x \leq 1.0$) and $Na_xCoO_2$ ($0.33 \leq x \leq 1.0$), have been a subject of an intense research activity in the past years owing to their potential technological applications as the battery electrodes and thermoelectric materials [1-3]. The notable structural and chemical features in this kind of materials are that the cation content can vary over a large range by deintercalation without evidently modifying the average crystallographic structure [4-6]. Recently, a rich variety of physical properties, such as superconductivity in $Na_{0.33}CoO_2 \cdot 1.3H_2O$ and charge ordering in $Na_{0.5}CoO_2$, have been observed [6-12]. Structural investigations revealed that the intercalations of either Na atoms or $H_2O$ molecules could make the local structure very complex; the intercalated atoms can be random with high mobility or crystallized in certain ordered states [12]. For instance, Na atoms in $Na_{0.5}CoO_2$ crystallize in a well-defined zigzag ordered pattern and yield an orthorhombic structure in which low temperature charge ordering is observed [8, 9, 12]. Intercalation of divalent ions ($Ca^{2+}$, $Sr^{2+}$) into cobalt oxides are expected to improve the thermoelectric properties. $Sr_{0.35}CoO_2$ shows a Seebeck coefficient comparable to that of $Na_xCoO_2$ [13]. Partial substitution of Ca for Na in $Na_xCoO_2$ also gave an enhancement in both Seebeck coefficient and electric resistivity [2, 14]. From structural point of view, significant questions to answer in $M_xCoO_2$ (M = Na, Li, Sr or Ca) materials are what kinds of ordered states actually exist for the intercalated M cations and whether the $CoO_2$ sheets show up relevant distortions depending on the M occupancy. Our recent experimental investigations on $Sr_xCoO_2$ ($0.25 \leq x \leq 0.4$) system demonstrated that the $Sr_{0.35}CoO_2$ material is crystallized in a well-defined Sr-ordered state yielding a superstructure of $3^{1/2}a \times 3^{1/2}a$. This specific superstructure is expected to have large effects on the charge ordered states and magnetic structures as theoretically demonstrated by first principle calculations [9, 18]. In this paper, we will focus our attention on superstructure modulations in correlation with Sr ordering and local structural distortion in $Sr_{0.35}CoO_2$. Certain fundamental physical properties and defect structures in present system have also been discussed.

Polycrystalline materials with nominal compositions of $Sr_xCoO_2$ ($0.25 \leq x \leq 0.4$) were prepared by the low-temperature ion exchange reaction from the $\gamma$-$Na_xCoO_2$ ($0.5 \leq x \leq 0.8$) precursor prepared by conventional solid-state reaction or by sodium deintercalation of $Na_{0.75}CoO_2$ [12]. The ion exchange process was carried out by using the modified Cushing-Wiley method [14]. A amount of $\gamma$-$Na_xCoO_2$ was mixed thoroughly with 10% molar excess anhydrous $Sr(NO_3)_2$ powder, then heated at 310°C for two days in air, the mixture was grinded repeatedly during the process. The final products are washed by distilled water. The metal ratios in the products were determined by inductively coupled plasma (ICP) analysis, yielding Sr : Co : Na = 0.36 : 1.0 : 0.0043 for $Sr_{0.35}CoO_2$ sample and 0.25 : 1.0 : 0.0035 for $Sr_{0.25}CoO_2$. Specimens for transmission electron microscopy (TEM) observations were polished mechanically with a Gatan polisher to a thickness of around 50 μm and then ion-milled by a Gatan-691 PIPS ion miller. The TEM investigations were performed on a TECNAI F20 operating at a voltage of 200 KV.

Fig.1 shows the XRD patterns obtained from materials with nominal compositions of $Sr_xCoO_2$ (x = 0.35 and 0.25) and their precursors $\gamma$-$Na_xCoO_2$ (x = 0.7 and 0.5). Structural analysis indicated that the average structures of both $Sr_{0.25}CoO_2$ and $Sr_{0.35}CoO_2$ materials are isomorphic to the $\gamma$-$Na_xCoO_2$ phase with the layered hexagonal structure. Lattice parameters calculated from our experiments are given in fig. 1. The evident change of lattice parameters is only observed along the c axis direction, which is considered arising from the relatively larger Sr substitution for Na among the $CoO_2$ layers. For instance, the $Sr_{0.35}CoO_2$ has the lattice parameter of c = 1.15 nm with a notable 7% increase in comparison with that of $Na_{0.7}CoO_2$ (c = 1.08 nm). The Structural transition from orthorhombic $Na_{0.5}CoO_2$ to hexagonal $Sr_{0.25}CoO_2$ can be also seen in Fig.1 (b). This fact directly suggests that the orthorhombic structural distortion in $Na_{0.5}CoO_2$ which arises from the zigzag Na ordering is totally destroyed in the $Sr_{0.25}CoO_2$ sample. It is also noted that the reflection peaks for $Sr_{0.25}CoO_2$ are somewhat broader than that of $Sr_{0.35}CoO_2$. Further TEM investigation reveals that $Sr_{0.35}CoO_2$ is crystallized in a well defined Sr ordered phase, while $Sr_{0.25}CoO_2$ in general contains notable structural inhomogeneity and defects.

Fig. 2 (a) shows the temperature dependences of resistivity of $Sr_{0.35}CoO_2$ annealed respectively at 300 °C in air and at 400 °C in oxygen for 8 hours. Our systematic analyses revealed that the resistivity behaviors of $Sr_xCoO_2$ materials depend sensitively on the annealing conditions. The resistivity curve for 300 °C annealed sample is characteristic of nonmetallic (semiconductor), i.e., $d\rho/dT<0$. On the other hand, the resistivity curve becomes metallic in the sample annealed at 400 °C in oxygen with $d\rho/dT>0$. Fig. 2 (b) shows the temperature dependences of magnetic susceptibility ($\chi$) at a field of 2 T for $Sr_{0.35}CoO_2$ annealed at 300 °C and 400 °C, respectively. For both the samples, $\chi$ increases with lowering temperature and obeys the Curie-Weiss law. At about 30 K, a small kink appears in both samples but more notable for 400 °C annealed sample. Above 50 K, $\chi$ can be well fitted by the formula $\chi(T)=C/(T-\theta)$, with $\theta$ = -133.1 K for 300 °C annealed sample and the 400 °C annealed sample -128.6 K (see the inset of Fig. 2 (b)). The negative value of $\theta$ indicates that the spin correlations are antiferromagnetic. The fitting parameters allow us to give rise to the effective moments as 1.36μB/Co and 1.38μB/Co for $Sr_{0.35}CoO_2$ annealed at 300 °C and 400 °C, respectively. The value of effective moments for $Sr_{0.35}CoO_2$ is close to that of $Na_xCoO_2$ with x ~0.7 [15]. Though the resistivity behavior in present system is sensitive to the annealing conditions, the magnetic properties is almost kept inflexible in the annealed samples, this fact suggests that the annealing processing may mainly change the oxygen content in the regions of grain boundaries which normally play an important role to understand the nature of resistivity.

In order to understand the microstructure features of the $Sr_{0.35}CoO_2$ materials, we have performed a series of investigations by means of scanning electron microscopy (SEM), selected-area electron diffraction and high-resolution transmission electron microscopy (HRTEM). Figs. 3 (a) and (b) show SEM images illustrating the typical morphological features of $Na_{0.70}CoO_2$ and $Sr_{0.35}CoO_2$ crystal grains. Both materials show clearly layered structure. Figs. 3 (c) and (d) show the electron-diffraction patterns for the $Sr_{0.35}CoO_2$ sample taken along the [001] and [1$\bar{1}$0] zone-axis directions, respectively. All main diffraction spots with strong intensity can be well indexed by a hexagonal unit cell with lattice parameters of a = 0.282 nm, and c = 1.152 nm and a space group of $P6_3/mmc$ with the reflection conditions of L=2n for (h h 0) and L=2n for (2h –h 0). The most notable structural phenomenon revealed in the electron diffraction patterns is the appearance of systematic weak reflection spots along the <110> direction. This superstructure can be well characterized by an in-plane wave vector $\mathbf{q}_1$ = (1/3, 1/3, 0) which yields a $3^{1/2}a \times 3^{1/2}a$ super-cell within the basic *a-b* plane. The superlattice spots on the **a\*-b\*** plane in general are very sharp, indicating a relatively long length (>30nm) of the ordered state. The superstructure reflections observed in the diffraction pattern along the [110] zone axis exhibit as weak diffuse spots that are extended along the **c\***-direction as shown in fig. 3 (d), indicating a short coherence length perpendicular to the $CoO_2$ planes (<5nm). This superstructure can be well interpreted by Sr ordering among $CoO_2$ sheets. Detailed structural analysis on this ordered state will be discussed with HRTEM investigations in following context.

Another notable structural feature is observed in the [010] zone-axis electron diffraction pattern, as shown in fig. 3 (e), where we can see a new series of superstructure reflection in addition to the main spots as indicated by arrows. Careful structural analysis indicates that this superstructure corresponds to a structural modulation with $q_2 = a^*/2$ resulting from atomic shifts along the c axis direction. This kind of structural distortion therefore is invisible in the TEM observation along [001] zone-axis as demonstrated in fig. 3 (c). The weak superstructure reflections as revealed in our observations normally appear as weak diffuse spots that are slightly extended along the $c^*$- direction. The coherence length perpendicular to the $CoO_2$ planes is estimated to be around 6 ~ 10 nm varying slowly from one area to another.

Systematic structural analysis on the $Sr_xCoO_2$ ($0.25 \leq x \leq 0.4$) materials suggests that the well-defined Sr-ordered state only appears at $x \cong 1/3$. Materials with other Sr concentrations contain complex structural imhomogeneities and phase separation. For instance, the diffraction pattern taken from a crystal in the sample with nominal composition of $Sr_{0.25}CoO_2$ shows the presence of complex diffraction streaks following with the $q_1$ superstructure spots as illustrated in fig. 3 (f).

A better and clear view of the atomic structure for the $q_1$ modulation is obtained along the c-axis direction. Fig. 4 (a) shows a [001] zone-axis HREM image for $Sr_{0.35}CoO_2$. This image was obtained from a thin region of a crystal under the defocus value around the Scherzer defocus (≈-60nm). The metal atom (Co) positions are therefore recognizable as dark dots. In the present case, the hexagonal superstructure is clearly seen in the relative thick area as illustrated by a hexagonal supercell. We therefore interpret this superstructure in terms of intercalated Sr-ordering as illustrated in fig. 4 (b). Previously, the structural nature of the intercalated Na ions in $Na_{0.7}CoO_2$ single crystal was studied by neutron diffraction in detail [8, 9]. This work showed that Na ions can occupy two possible positions ($P_1$ and $P_2$) governed by the space between the two adjacent O planes. Both of the sites are trigonal prismatic sites, but one trigonal prism shares edges with adjacent $CoO_6$ octahedral ($P_2$), while the other trigonal prism shares faces with adjacent $CoO_6$ octahedral ($P_1$). In present case, the $Sr^{2+}$ and negative charged $CoO_2$ layer interaction and $Sr^{2+}$ intra-plane screened electrostatic interaction is likely to favor the occupancy of Sr cation at the $P_2$ sites [13]. HRTEM image calculations based on this model were carried out by systematically varying the crystal thickness and defocus value. A calculated image for defocus value of –60 nm and a thickness of 3 nm, superimposed onto the experimental image, appears to be in good agreement with the experimental one as shown in fig. 4 (c)

Fig. 4 (d) shows a HRTEM micrograph directly showing the layered structural feature and the $q_2$ structural modulation along <100> direction. This image was obtained from a relative thick region (~10nm) to allow demonstration of the $q_2$ modulation. Essentially, the $q_2$-modulation is due to the development of a $CoO_2$ layer distortion depending on the Sr concentration and occupancy. In the $Sr_{0.35}CoO_2$ samples, only 33% percent of $P_2$ site are occupied, the local structural relaxation yields a notable atomic shifts mainly oriented along the *c*-axis, and which could form certain kinds of ordered state to minimize elastic energy, i.e. $q_2 = a^* / 2$ in present case. According to the kinematical theory for electron diffraction of distortion modulation, the transverse $q_2$-modulation can not result in any effects that are clearly observable in the TEM measurements along the *c*-axis direction [16]. This result is in good agreement with our observed results as demonstrated by the TEM results shown in Fig. 3 (c) and Fig. 4 (a). Another remarkable structural phenomenon noted in our study is that the periodicities of both $q_1$- and $q_2$-modulations in $Sr_xCoO_2$ are likely to be governed by commensurate effects, which are known to play an essential role in the study of the charge-ordering and cooperative John-Taller structural distortion in the high $T_c$ superconductors and manganites [17].

In conclusion, layered $Sr_xCoO_2$ ($0.25 \leq x \leq 0.4$) has been synthesized by means of an ion-exchange reaction from $Na_xCoO_2$ ($0.5 \leq x \leq 0.8$). Structural analysis suggests that the $Sr_xCoO_2$ materials have the hexagonal average structure similar to γ-$Na_xCoO_2$ (space group $P6_3/mmc$). The well crystallized $Sr_{0.35}CoO_2$ material can be either metallic or semiconducting depending on the annealing conditions and its magnetic susceptibility increases progressively with lowering temperature following with the Curie-Weiss law. Two superstructure modulations observed in $Sr_{0.35}CoO_2$ are considered as arising from the intercalated Sr-ordering and a periodic structural distortion, respectively. The periodicities of both modulations are likely to be governed by the commensurate effect giving the wave vectors $q_1 = a^*/3 + b^*/3$ and $q_2 = a^*/2$. Detailed electron diffraction analysis suggests that $q_2$ modulation is a typical transverse structural modulation with local structural polarization along the c-axis direction.

We would like to thank Miss Y. Li for the assistance in preparing TEM samples and Dr. R.I. Walton for fruitful discussions and help during manuscript preparation. The work reported here is supported by 'National Natural Foundation' and'Outstanding Youth Fund' of China.

Figure captions

Fig. 1 XRD patterns of $Sr_xCoO_2$ (x = 0.25, 0.35) and $Na_xCoO_2$ (x = 0.5, 0.7), indicating both $Sr_{0.25}CoO_2$ and $Sr_{0.35}CoO_2$ have layered hexagonal structures.

Fig. 2 The resistivity (a) and magnetic susceptibility $\chi$ (b) as a function of temperature for $Sr_{0.35}CoO_2$ samples annealed at different temperatures and atmospheres. Inset of Fig. 2 (b) shows the theoretical fitting following with Curie law.

Fig. 3 (a) and (b) SEM images show the layered structural features of $Na_{0.7}CoO_2$ and $Sr_{0.35}CoO_2$. (c) and (d) Electron diffraction patterns of $\mathbf{q}_1$-superstructure in $Sr_{0.36}CoO_2$ taken respectively along the [001] and [1$\bar{1}$0] zone axis directions. (e) Electron diffraction pattern showing the presence of the $\mathbf{q}_2$ superstructure taken along [010] zone axis direction in $Sr_{0.35}CoO_2$. (f) Electron diffraction pattern showing the presence of complex ordered state in $Sr_{0.25}CoO_2$.

Fig. 4 (a) the [001] zone-axis HRTEM image showing the hexagonal sublattice and $\mathbf{q}_1$-superstructure. (b) Structural model for $Sr_{0.35}CoO_2$. The $3^{1/2}\boldsymbol{a} \times 3^{1/2}\boldsymbol{a}$ Sr-ordering is apparently illustrated. (c) Experimental and theoretical HRTEM images demonstrating the $\mathbf{q}_1$-superstructure from Sr ordering. (d) HRTEM image showing the $\mathbf{q}_2$ superstructure along <100> direction. An antiphase boundary is indicated by arrows.

Figure 1

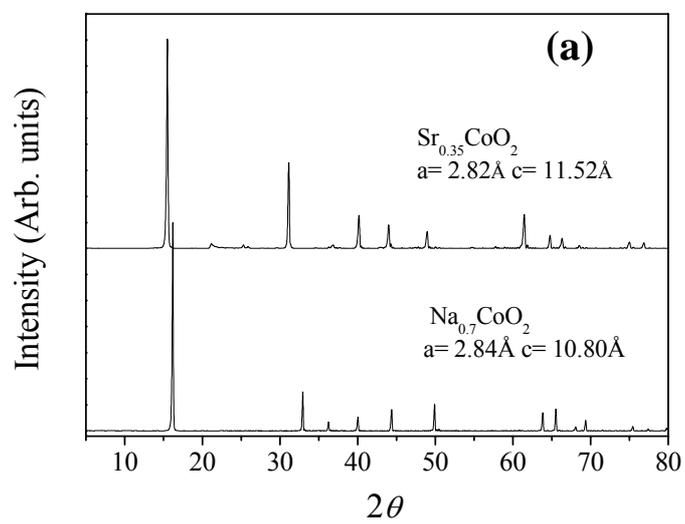

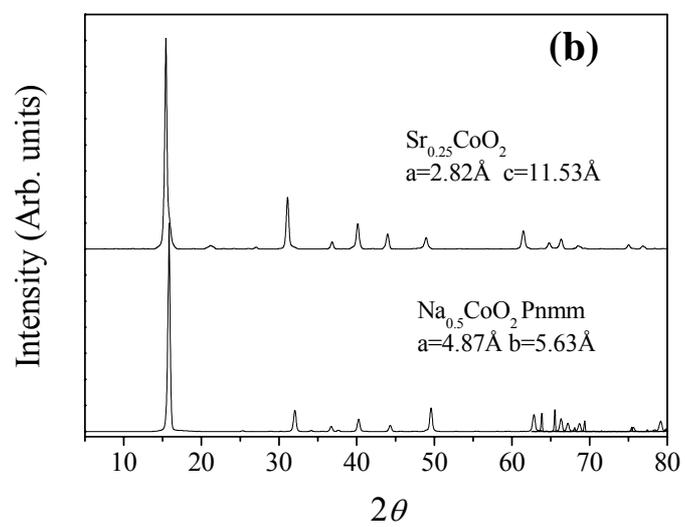

Figure 2

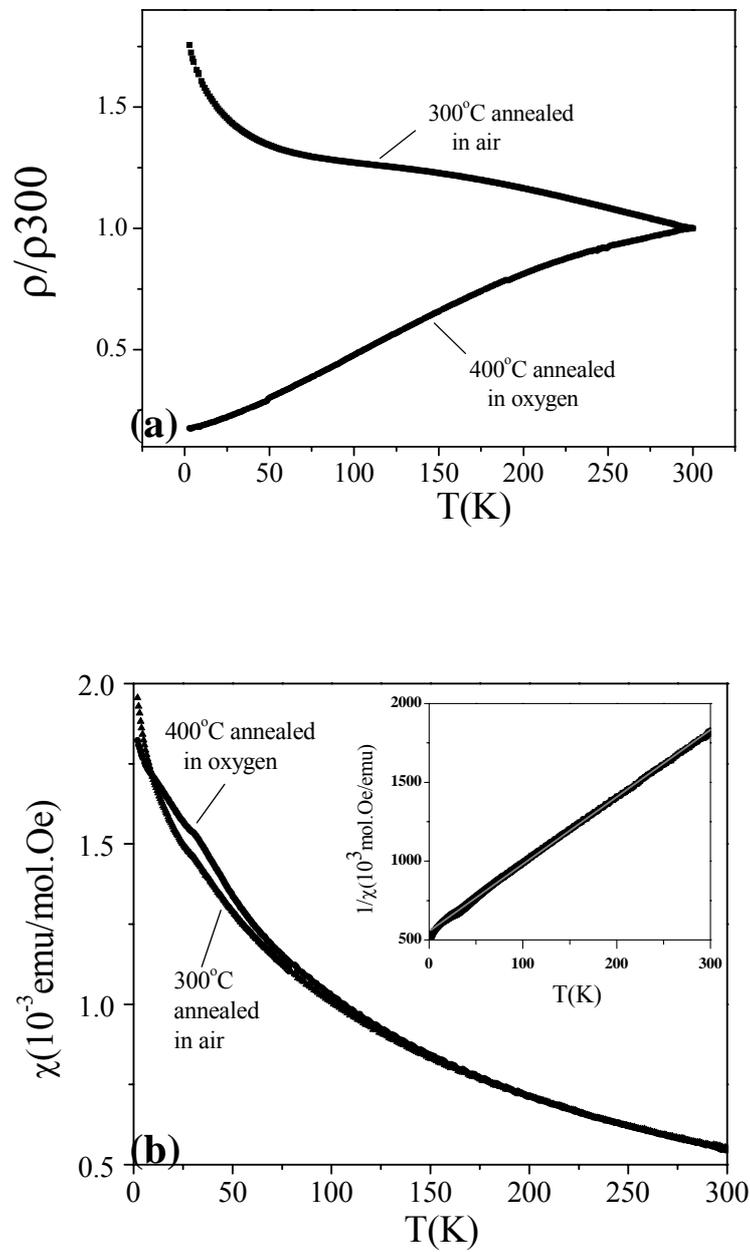

Figure 3

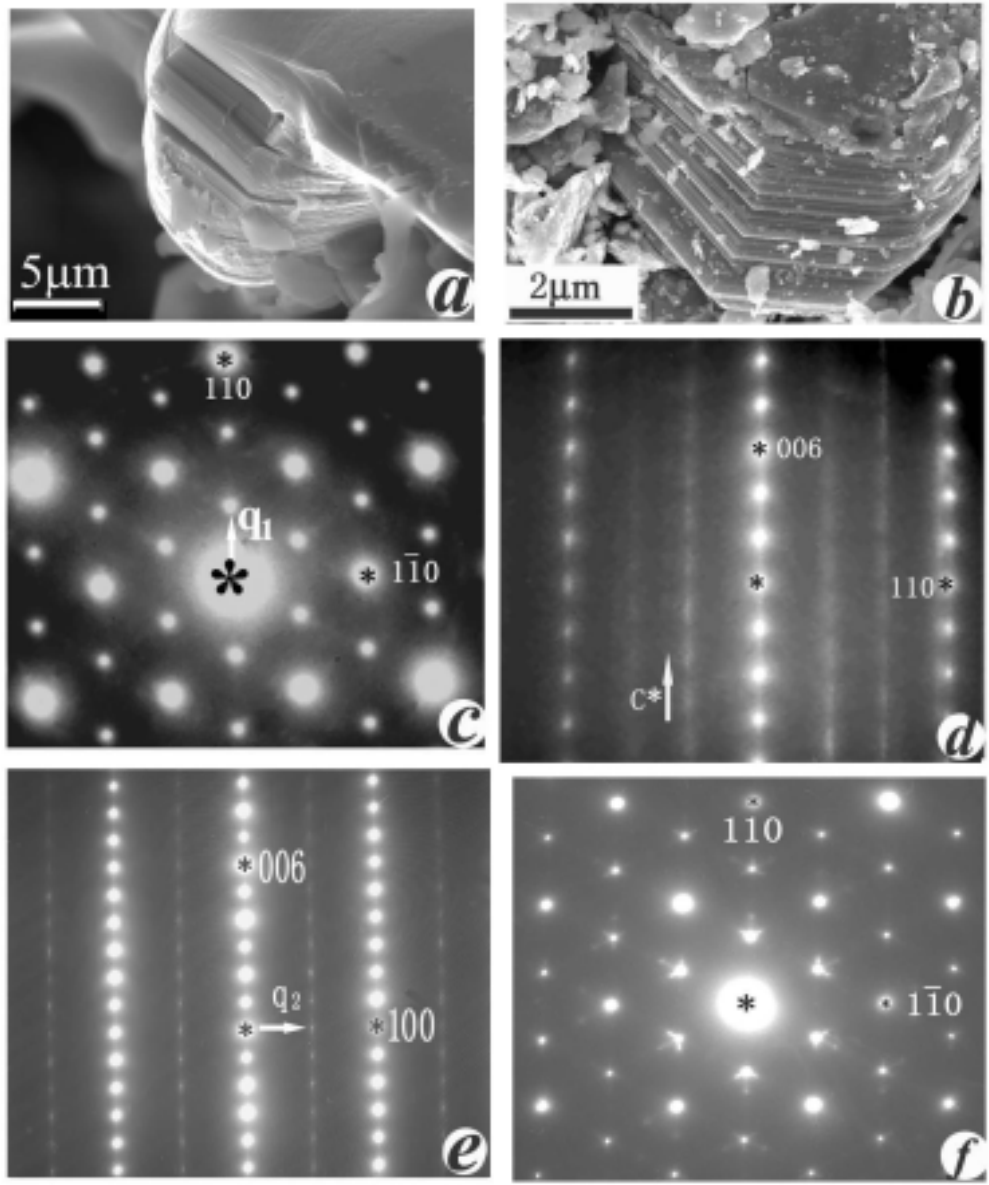

Figure 4

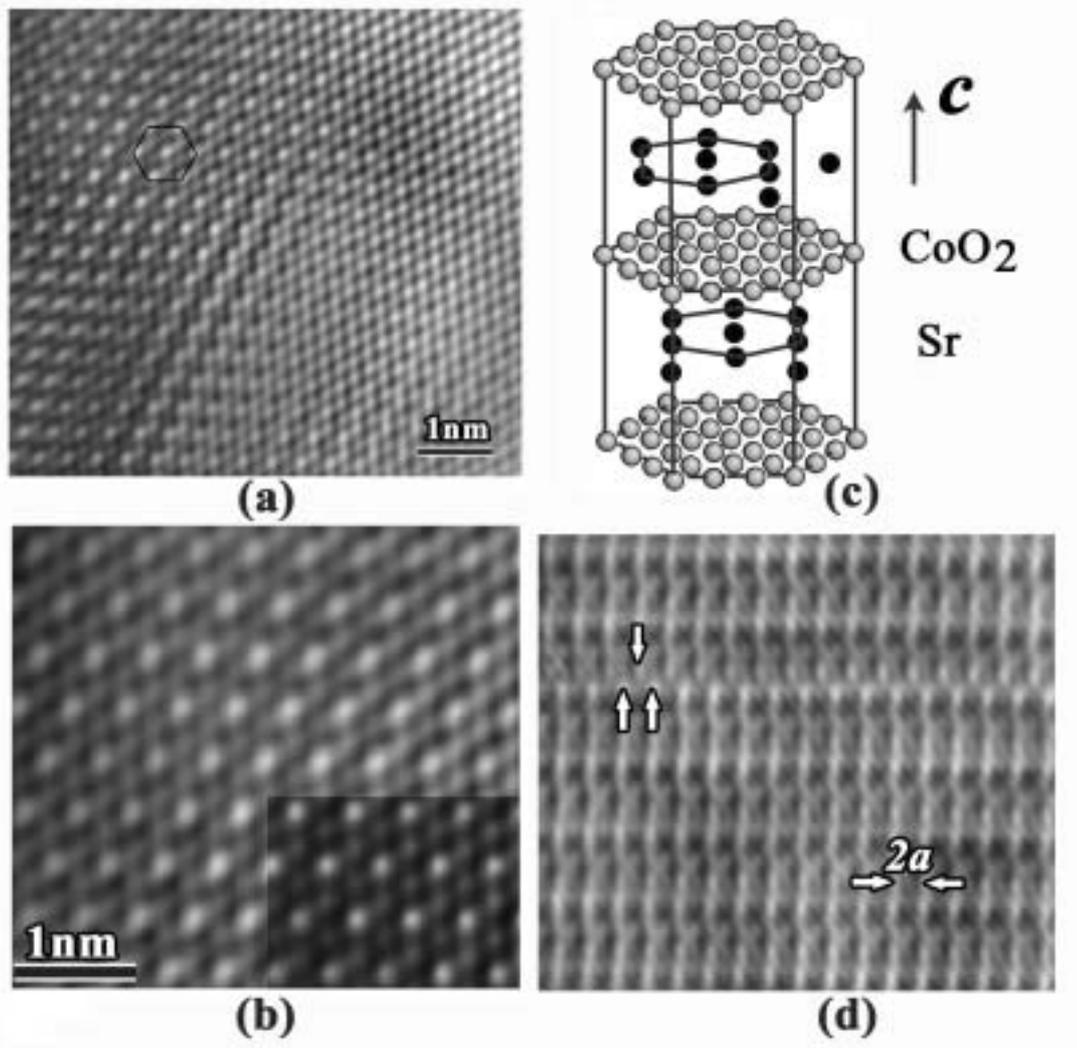